\documentclass[letterpaper,10.75pt]{article}
\usepackage[total={17cm,24cm},top={2cm},left={2.5cm}]{geometry}
\usepackage[utf8]{inputenc}
\usepackage{multicol,graphicx,anyfontsize,hyperref}
\usepackage{graphicx}
\usepackage{marginnote}
\usepackage{fancyhdr}
\usepackage[shortlabels]{enumitem}
\usepackage{amsfonts,amssymb,amsmath,wrapfig,lipsum}
\usepackage{float}
\usepackage{xcolor,cancel,comment,multirow}
\usepackage{pdfpages} 
\usepackage[font=footnotesize]{caption}
\usepackage{multicol}

\providecommand{\keywords}[1]
{\\
  \small\\
  \textbf{\textit{Key words---}} #1
}

\begin{document}

	\renewcommand{\refname}{Bibliography}
	\begin{flushleft}
		\scriptsize{ \scriptsize{UNIVERSIDAD AUTÓNOMA DE CHIHUAHUA}\hfill August - December 2021 Semester}
		\rule[.2cm]{\textwidth}{.001cm}
	\end{flushleft}
	\vspace{-.28cm}
	\begin{minipage}{3cm}
		\includegraphics[width=3cm]{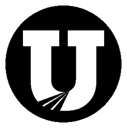}
	\end{minipage}\begin{minipage}{11cm}
		\centering
		\begin{tabular}{c}
			\resizebox{7cm}{!}{\Huge{\sl{\textbf{Detection of brain tumors }}}}\\
			
			    \resizebox{9.5cm}{!}{\Huge{\sl{\textbf{using machine learning algorithms}}}}\\
		
			\small{\tt{Pattern Recognition}}\\
		
			\small{\tt{M. I. Joseph Isaac Ramírez Hernández}}
		\end{tabular}
	\end{minipage}\begin{minipage}{3cm}
	
	\end{minipage}\\
	\rule{\textwidth}{.1cm}
	\begin{center}
		\normalsize{\textbf{H. Corral 320695, J. Melchor 320722, B. Sotelo 320711, J. Vera 324651  }}\\
			
				\normalsize{Circuito Universitario S/N, Nuevo Campus Universitario C.P. 31125, Chihuahua Chih. México. }\\
		\small{\sl{Facultad de Ingeniería}}\\
		
	\end{center}


\begin{abstract}
    \noindent
    An algorithm capable of processing NMR images was developed for analysis using machine learning techniques to detect the presence of brain tumors.
    
\end{abstract}

\begin{multicols}{2}

\section{Introduction}
Brain tumors rank 19th among all neoplasms, and 10th among the most lethal. In Mexico it is estimated that an incidence of 3.5 per 100,000 inhabitants and represents the second and fifth causes of cancer mortality in the age groups from 0 to 18 and 18 to 29 years.
\cite{presen1}.\

An early diagnosis is a determining factor in the survival chances of the patient. A severity 4 tumor, for example, is capable of doubling in size in 25 days, reducing the patient's life expectancy to less than 1 year \cite{presen2}. Among the techniques used for detection is Nuclear Magnetic Resonance (NMR), a non-radioactive and non-intrusive process that allows obtaining high-resolution images of the interior of the human body. Although it has limitations due to the fact that not all tissues are visible to the technique, MRI does allow the area of a brain affected by cancer to be clearly visualized, which gives to the technique reliability against this condition.\

In this way, the images can be used for the training of machine learning algorithms in order to create a useful tool for the doctors in charge of the diagnosis. Specifically, an algorithm capable of detecting whether or not a patient has a brain tumor can speed up and / or facilitate the diagnostic process by giving to the medical specialist an insight of the patient's situation.\

With this motivation in mind, a public bank of MRI images with confirmed tumors and another public bank of healthy patients were used to train, and subsequently compare, the performance of 4 classification algorithms: \textit{Decision Tree}, \textit{Random Forest}, \textit{AdaBoost} and \textit{Support Vector Machine}.

\section{Related work}

The work \textit{``A distinctive approach in brain tumor detection and classification using MRI''}, describes an automated system for the detection of brain tumors and lesions by MRI, with three steps of pre-processing, feature extraction and classification analysis. During pre-processing, different methods were applied to segment the region of interest. SVM with different kernels was used for classification; such kernels were Gaussian, linear, and cubic kernels.\

Another important work was "Brain tumor detection from MRI images using deep learning techniques", in which we found interesting that they worked with the same database as us. The techniques used were Artificial Neural Network (ANN) and Convolution Neural Network (CNN) where the precision of the training and validation data reaches 97.13 \% and 71.51 \%, and 89 \% and 94 \% , respectively. Despite using neural network methods, we believe is useful to know their work.

\section{Methodology}
\subsection{Pre-processing}
We began our work with a database of 253 brain magnetic resonance images of different dimensions, in which 155 of them were identified as resonances with the presence of tumors and 98 as healthy.\

First, the size of the images was adjusted to 300x300 pixels to facilitate their manipulation and processing. An image is nothing more than a matrix \textit{m$\times$n} where each pixel is an element of it, represented by a numerical value ranging from 0 to 255 (in grayscale). These values were captured so that each row was placed immediately after the previous one in a single row to form a matrix whose dimensions were reduced by \textit{m$\times$n} a 1$\times (n\cdot m)$ columns (characteristics). These rows were stored in another matrix, so that each row corresponded to an image, as shown in Figure \ref{DIAGRAM}. Thus forming our database. In particular, our database contains 253 lines and 90'001 columns, where the last column corresponds to a label with a 1 or a 0, depending on whether it is an image with or without a tumor, respectively.

\begin{center}
\includegraphics[width=1.0\linewidth]{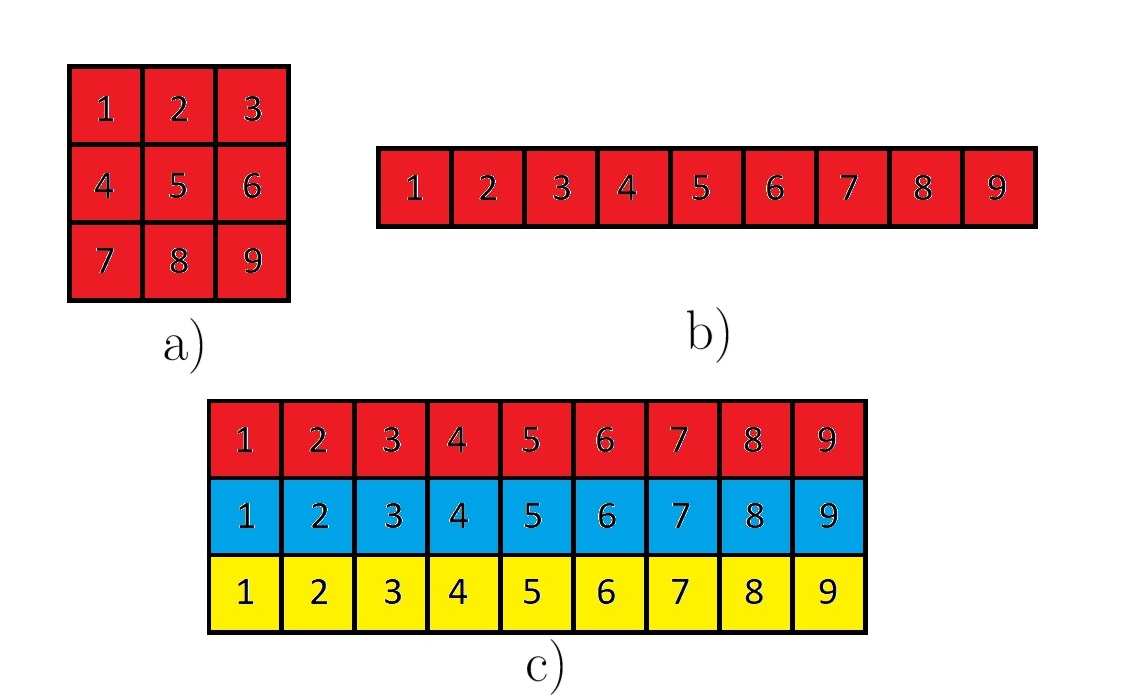}
\captionof{figure}{\textit{a)} An image is a matrix of \textit{n$\times$m} pixels where each pixel contains a numeric value. \textit{b)}Each line of the image was placed after the previous one in a single row. \textit{c)} These rows were arranged in a cascade to form a single matrix, which was used as our database.}
\label{DIAGRAM}
\end{center}

Subsequently, the values of the matrix were normalized and stored in a file .txt to be used in the next stage of the project.\

\subsection{Classification}

Once the preprocessing was finished, in another Python file we loaded the database and proceeded to separate the columns between characteristics (all rows and columns from 0 to 90,000) and label (column 90,000), storing them in separate variables (\textit{features} y \textit{targets}, respectively). Then, PCA was applied to identify the most significant characteristics in \textit{features} (of which the first 60 most significant ones were taken) and it was divided into a training set and a test set, using the library sklearn.model, so that the first set contained 80\% of the total data and the second the remaining 20\%. These sets were used to train the classifier algorithms.\ 

Each one of the chosen algorithms was executed with the instructions of print its percentage of precision and classify a test image (in which there is the presence of tumor). It is normal for the results to vary each time the code is executed, so the chosen strategy was to execute it 10 times and take an average of the precision percentages of each algorithm in the different categories and observe how often the classification was correct of the test image.\

In the particular case of SVM, different combinations of parameters were tested until the combination that produced the best predictions was found, and then the code was executed repeatedly as mentioned above. \

The following section shows the results obtained with each algorithm.

\section{Results}

The relevant categories were: model accuracy, referring to the percentage of elements of the test base that the algorithm classified correctly; percentage of sick, which is understood as the performance of the trained algorithm when it is subjected to classify the elements labeled as sick; and percentage of not sick, similar to the previous category, but with the items labeled as not sick. The percentage of test refers to the precision of the algorithm to classify an NMR image with tumor, foreign to our dataset.  \

\end{multicols}

\begin{table}[h!]
    \centering
    \begin{tabular}{|c|c|c|c|c|}
    \hline
        Algorithm & Model Accuracy (\%)& P. sick (\%) & P. not sick (\%) & P. Test(\%) \\ \hline
        \hline
        Decision Tree & 72.54 & 79.34 & 64.39&30  \\ \hline
        Random Forest & 78.43 & 92.08 & 58.53&50  \\ \hline
        Adaboost & 75.88 & 81.77 & 66.94&70  \\ 
 \hline
    \end{tabular}
    \caption{Results obtained. The table shows the precision percentages of the different algorithms.}
    \label{tab:my_label}
\end{table}

\begin{table}[h!]
    \centering
    \begin{tabular}{|c|c|c|c|}
    \hline
        \textbf{Kernel} & \textbf{C} & \textbf{Gamma} & \textbf{Degree} \\ \hline
        linear, sigmoid, rbf, polynomial & 0.1, 1, 2, 3, 4 & auto, scale & 2, 3, 4, 5 \\ \hline
    \end{tabular}
    \caption{Tested parameters in SVM algorithm.}
    \label{tab:my_label}
\end{table}

\begin{table}[h!]
    \centering
    \begin{tabular}{|c|c|c|c|c|c|}
    \hline
        \textbf{Kernel} & \textbf{C} & \textbf{Gamma} & \textbf{Degree} & \textbf{Accuracy} (\%)& \textbf{P. Test}(\%)\\ \hline
        rbf& 4 & scale & 2 & 74.68 & 100\\ \hline
    \end{tabular}
    \caption{Best combination of SVM.}
    \label{tab:my_label}
\end{table}

\begin{multicols}{2}

None of the first three algorithms stood out especially in their performance. While Random Forest had a high percentage of sick, it also had a low percentage of healthy and test image. On the other hand, in the AdaBoost the precision was slightly lower than in Random Forest, but it was compensated by a smaller difference between the percentages of sick and non-sick. AdaBoost's test percentage was considerably better than Decision Tree and Random Forest.\

On the other hand, Suport Vector Machine correctly classified 100 \% of the cases in the test image and had acceptable model accuracy. However, the comparison is incomplete because the percentages of sick and healthy were not included.\ 

None of the algorithms required excessive time to execute, however, this could be due to the size of our dataset. It is possible that with a higher one, the computational costs increase, as well as the processing time.

\section{Conclusions and future work}

It was possible to develop 4 algorithms capable of classifying NMR images, with different percentages of success and room for improvement. \

It was observed that the Suport Vector Machine performed well, although further analysis is required with this technique to have a better comparison between algorithms. \

An efficient algorithm was developed for the pre-processing of images to be analyzed by classification algorithms. \

The possibility of carrying out future work using a greater number of images for algorithm training and / or images with higher resolution is opened. Likewise, the objectives and results of this work can be taken to the field of neural networks, where there is extensive related work. \

\renewcommand{\refname}{References}

    \nocite{*}
	\bibliographystyle{IEEEtran}
	\bibliography{bib}

\end{multicols}

\end{document}